# Performance Analysis of QoS in PMP Mode WiMax Networks

Harwinder singh
Department of Computer Science, Punjabi University, Patiala
harwinder002@gmail.com

Maninder Singh
Department of Computer Science, Punjabi University, Patiala
singhmaninder25@yahoo.com

*Abstract* -**IEEE 802.16 standard supports two different topologies: point to multipoint (PMP) and Mesh. In this paper, a QoS mechanism for point to multipoint of IEEE 802.16 and BS scheduler for PMP Mode is proposed. This paper also describes quality of service over WiMAX networks. Average WiMAX delay, Average WiMAX load and Average WiMAX throughput at base station is analyzed and compared by applying different scheduler at Base station and at fixed nodes.**

## 1. INTRODUCTION

IEEE 802.16 is a set of telecommunications technology standards aimed at providing wireless access over long distances in variety of ways- from point to point to full mobile cell type access. IEEE 802.16 standard is developed to serve fixed subscriber stations (SSs) through a central base station (BS) using a PMP topology. In PMP mode, every subscriber stations are directly communicate with central base station. PMP mode(in WiMAX) easily provide different type of services than wired networks at lower cost of arrangement.

IEEE 802.16 is developed with QoS in mind. In PMP mode, five different service classes are introduced for different application and packets from different service classes are handled based on their QoS constraints. In this paper, QoS mechanism using WFQ queue compare with DWRR queue in PMP mode(in WiMAX).

## 2. PMP MODE OF IEEE 802.16

IEEE 802.16 -2004 defined in 2004, operates in 2-11 GHz as well as the original 10-66 GHz band ,provides medium data rates and supports PTP and PMP operation modes for fixed subscribers only. Only LOS and NLOS communication are supported. Where communication made possible between transmitter and receiver(s) are placed on high rise towers so as to avoid all physical obstacles between them, is called *line-of-sight(LOS)* and when LOS communication is not possible(e.g. when transmitter/receivers are devices inside a home), signals transmitted from the receiver undergo attenuation and multipath distortion (after bouncing off trees and building ). This type of communication is called *non line-of-sight (NLOS)* communication.

In PMP mode, Physical and Medium Access Control Layer plays important role in communication between base station and subscriber stations. WiMAX defines the concept of *service flow*. A *service flow* is unidirectional flow of packets with a particular set of *quality of service (QoS)* parameters. A service flow is identified by a 32-bit service flow identifier (SFID). In [1], N. Srinath describes WiMAX is connection-oriented protocol. This connection-oriented scheme provides a means for handling bandwidth requests and allocation traffic and QoS parameter with service flow etc. A connection is identified by a 16-bit connection identifier (CID).

MAC layer divide into three sub layers, first *service specific convergence sublayer (CS)* define interface with higher layers, converts higher layer packets into MAC level service flow and parameters. Second, MAC *Common Part sublayer (MAC CPS)* implements common MAC functionalities like link initialization, admission control, controlling channel access, transmission scheduling, quality of service, fragmentation, error control and retransmission. Third, *Security Sublayer* provides security through authentication, key management and encryption.

### 2.1 IEEE 802.16 MAC PROTOCOL

PMP architecture, which consists of one BS managing multiple SSs. Transmissions between the BS and SSs are realized in fixed-sized frames by means of time division multiple access (TDMA) / time division duplexing (TDD) mode of operation . According to Alexey Vinel[2], the frame structure consists of downlink sub-frame for transmission from the BS to SSs and an uplink sub-frame for transmission in the reverse direction as shown in fig 1. The Tx/Rx transition gap (TTG) and Rx/Tx transition gap(RTG) shall be inserted between the sub-frames to allow terminals to turn around from reception to transmission and vice versa. In the downlink sub-frame the Downlink Map (DL-MAP) and Uplink Map (UL-MAP) message are transmitted by the BS, which comprise the bandwidth allocation for data transmission in both downlink and uplink direction, respectively.

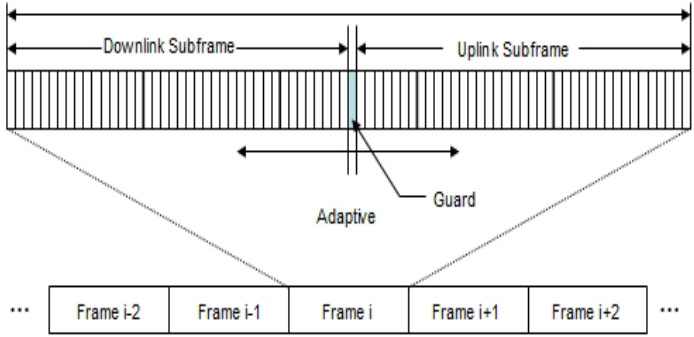

Figure1. TDMA-Frame structure

**2.2 IEEE802.16 QoS Classes and Scheduling**

IEEE 802.16 standard can support multiple communication services (data, voice and video) with different QoS requirements. The MAC layer defines QoS signaling mechanisms and functions that can control BS and SS data transmissions.

On the downlink, the transmission is relatively simple, because the BS is the only one that transmit during a downlink sub-frame. Data packets broadcast to all SSs and an SS only listens in on the the packets destined for it. On the uplink, the BS determines the number of time slots for which each SS will be allowed to transmit in an uplink sub-frame. This information is broadcast by the BS through the uplink map message (UP-MAP) at the beginning of each frame. The UL-MAP contains an information element (IE) per SS, which includes the transmission opportunities for each SS, i.e., the time slots in which a SS can transmit during the uplink sub-frame. The BS uplink-scheduling module determines the IEs by using the bandwidth request message sent from the SSs to the BS.

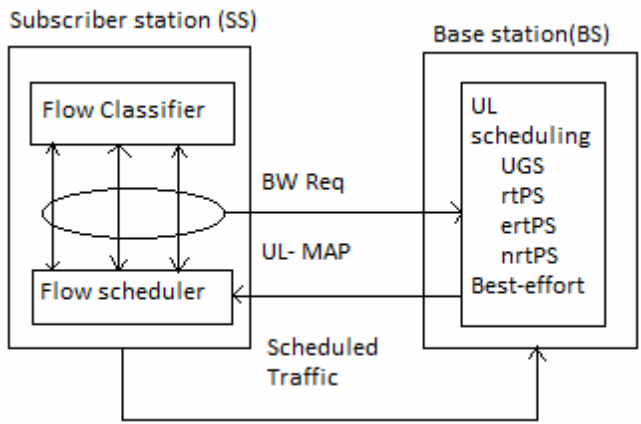

Figure 2. QoS Architecture of IEEE802.16

In the IEEE 802.16 standard, bandwidth-requests are normally transmitted in two modes according to Hemant kumar rath in [3]: a *contention mode* and *contention-free mode*(polling). In the contention mode, the SSs send bandwidth-requests during a contention periods, and contention is resolved by the BS using exponential back-off strategy. In the contention-free mode, the BS polls each SS, and an SS in reply sends its BW-request. There are five types of basic services described in the standard. Namely, Unsolicited Grant Service (UGS); Real-Time Polling Service (rtPS); Non-Real-Time Polling Service (nrtPS); Extended-Real-Time Polling Service (ertPS); Best-Effort (BE) service. Variable bandwidth assignment is possible in rtPS, nrtPS, ertPS and BE services. Whereas UGS service needs fixed and dedicated bandwidth assignment. Figure 2 shows the QoS architecture of IEEE 802.16 based services.

UGS is designed for constant bit-rate (CBR) like flows such as VoIP which require constant bandwidth allocation. rtPS service is designed for variable bit-rate (VBR) flows such as MPEG video, which have specific bandwidth requirements as well as the latency. ertPS builds on the efficiency of both UGS and rtPS and designed to support real-time service flows that generate variable-size data packets on periodic basis, such as voice over IP services with silence suppression. According to Yanqun Le,Yi Wu[4], the nrtPS and BE are for VBR non-real time applications(e.g. bandwidth intensive file transfer) and best-effort applications(e.g. HTTP), respectively.

In [5],Aun Haider and Richard j. Harris describes packets schedulers can be classified into the following two types: work conserving and nonwork conserving . Examples of work conserving scheduling algorithms include Generalized Processor Sharing (GPS), Weighted Round Robin (WRR), Deficit Weighted Round Robin (DWRR),Weighted Fair Queueing (WFQ), Self Clocked Fair Queueing (SCFQ); whereas Hierarchical Round Robin (HRR), Stop-and-Go, and Jitter-Earliest-Due-Date are some examples of nonwork conserving schedulers.

In our Proposed QoS mechanism, we have used Deficit Weighted Round Robin (DWRR) and Weighted Fair Queuing (WFQ) schedulers. In DWRR is modified weight round robin scheduling discipline. It can handle packets of variable size without knowing their mean size. A maximum packet size number is subtracted from the packet length and packets that exceed that number are held back until the next visit of the scheduler. WRR serves every non empty queue whereas DWRR serves packet at head of every non-empty queue whose deficit counter is greater than the packet's size at Head of Queue (HoQ) if the deficit counter is lower, then the queue is skipped (HoQ packet is not served) and its credit increased by some given value called quantum. The increased value is used to calculate the deficit counter the next time around when the scheduler examines this queue for serving its head-of-line. If the queue is served, then the credit is determined by the size of packet being served. In [5], Aun Haider describes that DWRR is simple O(1). It can be employed for scheduling at the BS of a WiMAX network.

Weighted Fair Queueing (WFQ) is data packet scheduling technique allowing priorities statistically multiplexed data flows WFQ is a generalization of fair queuing (FQ). Both in WFQ and FQ, each data flow has a separate FIFO queue. In FQ, with a link data rate of $R$, at any given time the $N$ active data flows (the ones with non-empty queues) are serviced simultaneously, each at an average data rate of $R/N$. Since each data flow has its own queue, an ill-behaved flow (who has sent larger packets or more packets per second than the others since it became active) will only punish itself and not other sessions. Contrary to FQ, WFQ allows different sessions to have different service shares. If $N$ data flows currently are active, with weights $w_1, w_2...w_N$, data flow number $i$ will achieve an average data rate of

$$\frac{Rw_i}{(w_1+w_2+w_3+\ldots\ldots\ldots+w_N)}$$

Network with WFQ switches and a data flow that is leaky bucket constrained, an end-to-end delay bound can be guaranteed. By regulating the WFQ weights dynamically, WFQ can be utilized for controlling the quality of service.

## 3. PROPOSED QoS MECHANISM

The network topology of simulation scenarios is illustrated in Figure 3. There is one BS, five fixed nodes. We have applied DWRR and WFQ scheduler at each fixed node and at the base station but one can also use DWRR scheduler at the BS (Base station) for scheduling in WiMAX network and WFQ Scheduler at fixed stations for scheduling the traffic belonging to the nrtPS class. But in our QoS mechanism, we have used DWRR and WFQ scheduler for five different traffic classes like UGS, rtPS, nrtPS, Best-Effort and ertPS at one BS and five fixed SSs in network topology. First we have assigned different interface having different IP addresses to the BS and five fixed nodes also called subscriber stations (SS) then applied DWRR and WFQ scheduler at BS and SSs and used best-effort type of service (TOS) respectively.

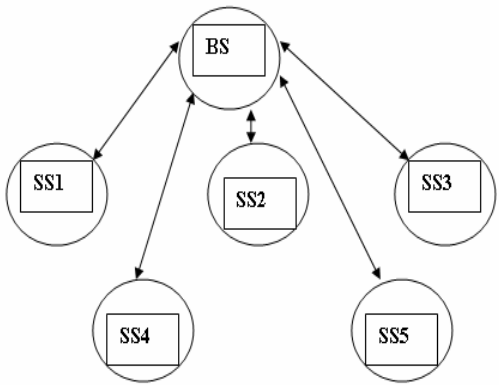

Figure 3. Network Topology

## 4. SIMULATION SCENARIO

In the simulation, we have used a topology that consists of one base station (BS) and five fixed node (SSs). SS1 sends ftp traffic to SS2, SS2 sends video traffic to SS3, SS3 sends http traffic to SS4, SS4 sends VoIP with silence suppression and SS4 sends voice traffic to SS1 fixed node. We have assumed error free link conditions. Wireless OFDMA PHY layer of IEEE 802.16 standard is used with a channel bandwidth of 20MHz. The frame duration is 12.5 ms is used. ARQ and packing mechanisms are not used. Other simulation are parameters are provided in Table 1.

| Simulation parameter | Value |
|---|---|
| Channel Bandwidth | 20MHz |
| Frame Duration | 12.5 ms |
| TTG | 106 ms |
| RTG | 60 ms |
| Modulation scheme | 64 QAM,16 QAM |
| Coding rate | 3/4 |
| Duplexing Technique | TDD |

Table 1. Simulation Parameters

## 5. SIIMULATION RESULTS

To present the results of simulations we have compared Average WiMAX delay at base station (BS) and at each fixed node (SSs) using DWRR and WFQ scheduler with different type-of-service (TOS) respectively. In following simulation result figures, nnn-scenario1-DES-1 is referred to simulation run with WFQ scheduler and rrr-scenario1- DES-1 is referred to simulation run with DWRR scheduler.

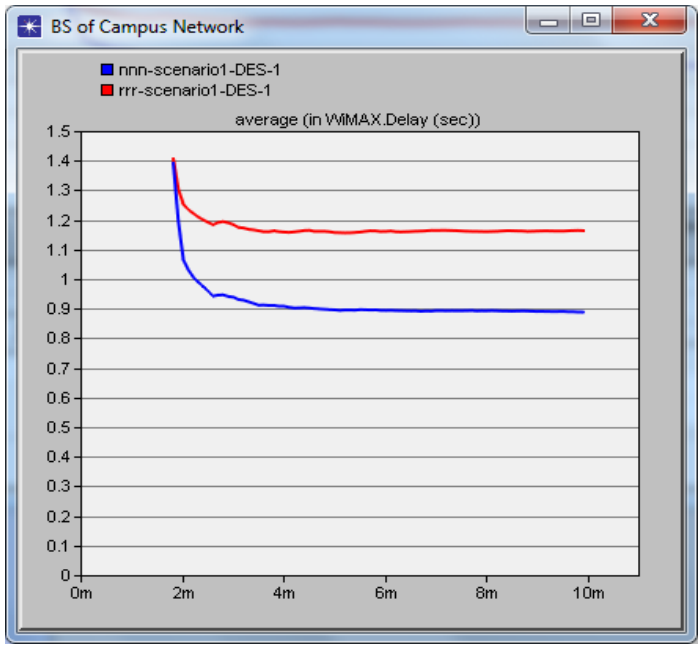

Figure 4 Average delay (in sec) in WiMAX .

Average WiMAX delay (in sec) using WFQ scheduler at base station is less as compared to average WiMAX delay using DWRR scheduler as shown in the fig 4. Average WiMax throughput (bits/sec) using WFQ scheduler at base station is higher than average WiMAX throughput (bits/sec) using DWRR scheduler at base station, shown in fig. 5.

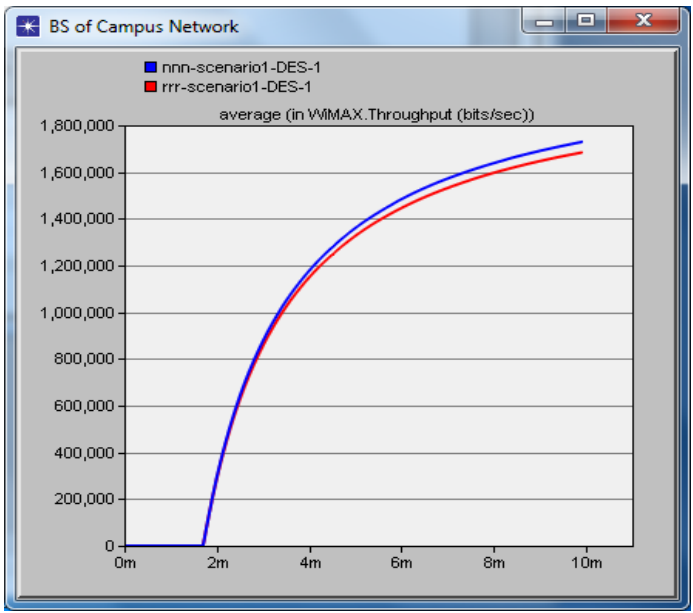

Figure 5. Average Throughput (bits/sec) in WiMAX .

But Average Load in WiMAX (bits/sec) is totally conversed to Average delay in WiMAX (in sec) at base station. i.e. Average Load in WiMAX (bits/sec) at base station using WFQ scheduler is higher than Average load in WiMAX (bits/sec) at base station using DWRR scheduler which is shown in fig 6.

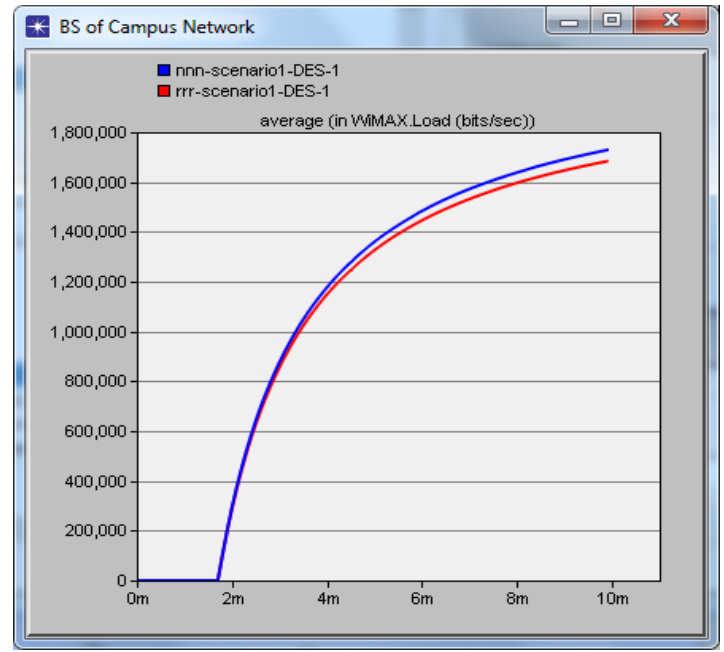

Figure 6. Average Load (in bits/sec) in WiMAX.

Next, we have shown the performance of interfaces which is used at base station using DWRR and WFQ scheduler. Traffic received/sent is higher at the base station if we are using Weighted Fair Queue (WFQ) scheduler as compared to the DWRR scheduler, as shown in fig. 7 & 8.

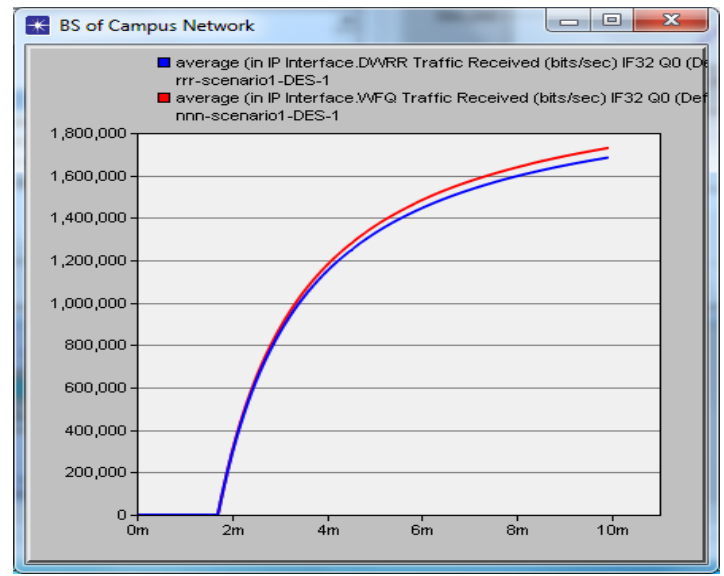

Figure 7. Traffic Received (bits/sec) through IP interface.

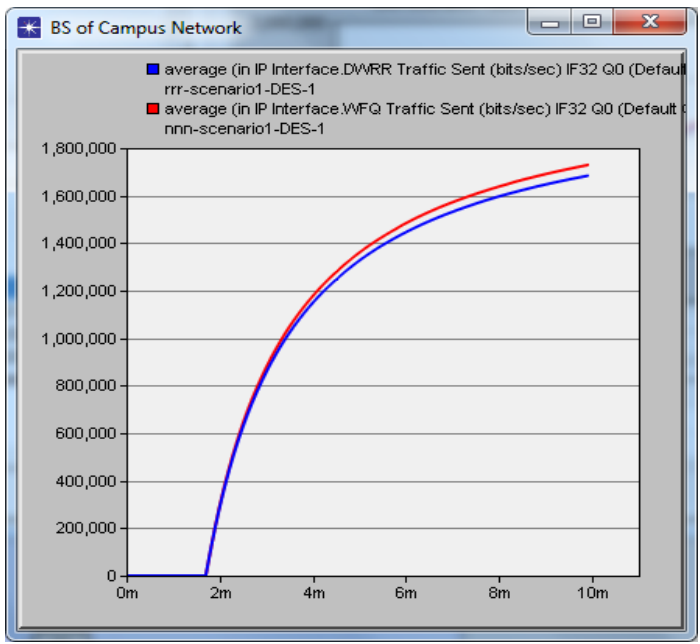

Figure 8. Traffic sent (bits/sec) through IP interface.

**6. CONCLUSION**

In this paper, we have proposed a QoS mechanism for WiMAX delay in PMP mode of IEEE 802.16. We have used simple scheduling for the base station and fixed nodes. In which WFQ performs better than DWRR scheduler. The results of the comparison have shown that IP interface gives better output for received and sent the traffic(bits/sec) and Delay in DWRR is more compared to using WFQ. The data transfer rate of DWRR is also less than WFQ.